\documentclass[twocolumn,prl,aps,showpacs]{revtex4-1}
\usepackage{graphicx}
\usepackage{color}
\usepackage{dcolumn}
\usepackage{amsmath}

\newlength{\figwidth} 
 \newlength{\figwidthb} %
\newcommand{\SIO}{Sr$_2$IrO$_4$ }
\newcommand{\SIOns}{Sr$_2$IrO$_4$}
\newcommand{\LCO}{La$_2$CuO$_4$ } 
\newcommand{\LCOns}{La$_2$CuO$_4$} 
\newcommand{\jeff}{$J_{\textrm{eff}}$}

\begin{document}

\title{Isospin Dynamics in \SIO: Forging Links to Cuprate Superconductivity}

\author{Jungho Kim,$^1$ D. Casa,$^1$ M. H. Upton,$^1$ T. Gog,$^1$ Young-June Kim,$^2$ J. F. Mitchell,$^3$ M. van Veenendaal,$^{1,4}$ M. Daghofer,$^5$ J. van den Brink,$^5$ G. Khaliullin,$^6$ B. J. Kim$^3$}

\address{$^1$Advanced Photon Source, Argonne National Laboratory, Argonne, Illinois 60439, USA}
\address{$^2$Department of Physics, University of Toronto, Toronto, Ontario, Canada M5S 1A7}
\address{$^3$Material Science Division, Argonne National Laboratory, Argonne, IL 60439, USA}
\address{$^4$Department of Physics, Northern Illinois University, De Kalb, IL 60115, USA}
\address{$^5$Institute for Theoretical Solid Sate Physics, IFW Dresden, Helmholtzstr. 20, 01069 Dresden, Germany}
\address{$^6$Max Planck Institute for Solid State Research, Heisenbergstraße 1, D-70569 Stuttgart, Germany}

\date{\today}

\begin{abstract}
We used resonant inelastic x-ray scattering to reveal the nature of
magnetic interactions in \SIOns, a 5$d$ transition-metal oxide with a
spin-orbit entangled ground state and \jeff=1/2 magnetic momemts, 
referred to as `isospins'. The magnon dispersion in \SIO is well
described by an antiferromagnetic Heisenberg model with isospin
one-half moments on a square lattice, which renders the low-energy
effective physics of \SIO much akin to that in superconducting
cuprates. This is further supported by the observation of exciton modes
in \SIO  whose dispersion is strongly renormalized by magnons, which
can be understood by analogy to the hole propagation in the background
of antiferromagnetically ordered spins in the
cuprates.
\end{abstract}


\maketitle

Quantum magnetism in transition-metal oxides (TMOs) arises from
superexchange interactions of spin moments that depend on spin-orbital
configurations in the ground and excited states. The array of
magnetism in 3$d$ TMOs are now well understood within the framework of
Goodenough-Kanamori-Anderson~\cite{GoodEnough}, which assumes conservation of spin
angular momentum in the virtual charge fluctuations. However, it has been recently realized that strong relativistic spin-orbit
coupling (SOC) can drastically modify the magnetic interactions and
yield far richer spectrum of magnetic systems beyond the standard
picture. Such is the case in 5$d$ TMOs, in which the energy scale of SOC
is on the order of 0.5 eV (as compared to $\sim$10 meV in 3$d$
TMOs). For example, A$_2$IrO$_3$ (A=Li,Na) is being discussed as a possible
realization of the long-sought Kitaev model with bond-dependent spin
interactions~\cite{jackeli09,chaloupka10,HillPRB11}. Furthermore, strong SOC may result in nontrivial band
topology to realize exotic topological states of matter with broken time reversal symmetry, such as a topological Mott
insulator~\cite{BalentsNPhys10}, Weyl semi-metal and axion insulator~\cite{WanPRB11}. Despite such
intriguing proposals abound, the nature of magnetic interactions in
systems with strong SOC remains experimentally an open question.

In this Letter, we report on the magnetic interactions in a 5$d$ TMO
\SIO with spin-orbit entangled ground state carrying \jeff=1/2
moments~\cite{bjkim08,bjkim09}, probed by resonant inelastic x-ray scattering (RIXS). We
refer to these \jeff=1/2 moments as `isospins.' On theoretical grounds,
magnetic interactions of isospins are expected to strongly depend on
lattice and bonding geometries. In the particular case of corner
sharing oxygen octahedra on a square lattice, relevant to
\SIO \cite{crawford94}(Fig. 1(a)), it is predicted that the magnetic interactions of isospins are
described by a pure Heisenberg model, barring Hund's coupling that
contributes a weak dipolar-like anisotropy term~\cite{jackeli09, wang11}. This is surprising considering
that strong SOC typically results in anisotropic magnetic couplings 
that deviate
from the pure Heisenberg-like spin 
interaction in the weak SOC limit. A
compelling outcome is that a novel Heisenberg antiferromagnet can be
realized in the strong SOC limit, on which a novel platform for high
temperature superconductivity (HTSC) may be designed.

RIXS in the last few years has become a powerful tool to study
magnetic excitations~\cite{ament11}. We report measurement of single magnons
using hard x-rays, which has complementary advantages over soft x-rays as detailed later on. The RIXS measurements were performed using the spectrometer at the 9-1D and the MERIX spectrometer at the 30-ID beamline of the Advanced Photon Source. A horizontal scattering geometry was used with the $\pi$ incident photon polarization. A spherical diced Si(844) analyzer was used. The overall energy and momentum resolution of the RIXS spectrometer at the Ir L$_3$ edge ($\approx$11.2 keV) was $\approx$130 meV and $\pm$0.032 $\AA^{-1}$, respecitvely.

\begin{figure}[t]
\hspace*{-0.1cm}\vspace*{-0.2cm}\centerline{\includegraphics[width=0.7\columnwidth,angle=0]{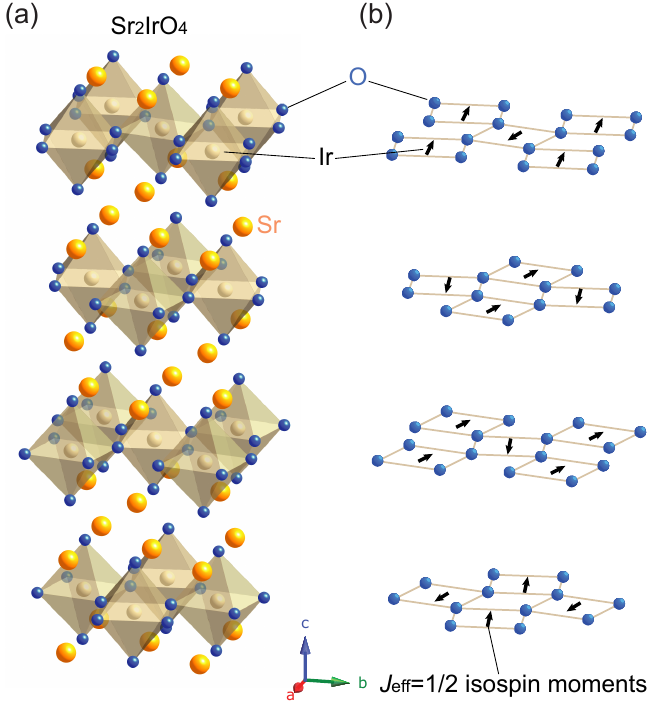}}
\vspace*{-0.3cm}%
%
\caption{(a) Due to a staggered in-layer rotation of oxygen octahedra, \SIO has four IrO$_2$ layers in the unit cell~\cite{crawford94}, which coincides with the magnetic unit cell. (b) Isospin one-half (\jeff=1/2) moments lie and are canted in the IrO$_2$ plane~\cite{bjkim09}.}\label{fig:fig1}
\end{figure}

\begin{figure}[t]
\hspace*{-0.2cm}\vspace*{-0.1cm}\centerline{\includegraphics[width=0.65\columnwidth,angle=0]{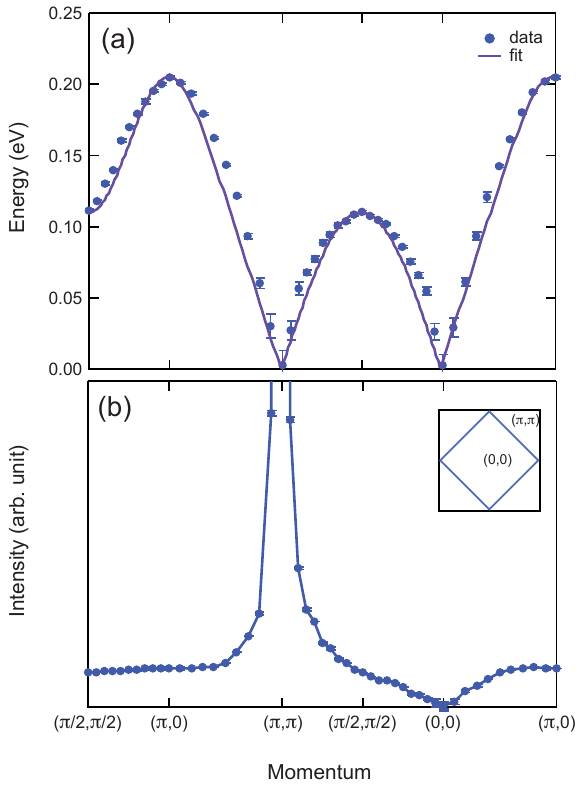}}
\vspace*{-0.5cm}%
%
\caption{(a) Blue dots with error bars show the single magnon dispersion extracted by fitting the energy loss curves shown in Fig.~\ref{fig:fig3}(b) ~\cite{supp}. The magnons disperse up to $\approx$205 meV at ($\pi$,0) and ≈110 meV at ($\pi$/2,$\pi$/2). Solid purple line shows the best fit to the data with $J$=60, $J'$=-20, $J''$=15 meV. (b) Momentum dependence of the intensities showing diverging intensity at ($\pi$, $\pi$) and vanishing intensity at (0,0). Inset shows the Brillouin zone of the undistorted tetragonal (I4/mmm) unit cell (black square) and the magnetic cell (blue square), and the notation follows the convention for the tetragonal unit cell, as, for instance, in \LCO.}\label{fig:fig2}
\end{figure}

\begin{figure*}[t]
\vspace*{-0.2cm}\centerline{\includegraphics[width=1.6\columnwidth,angle=0]{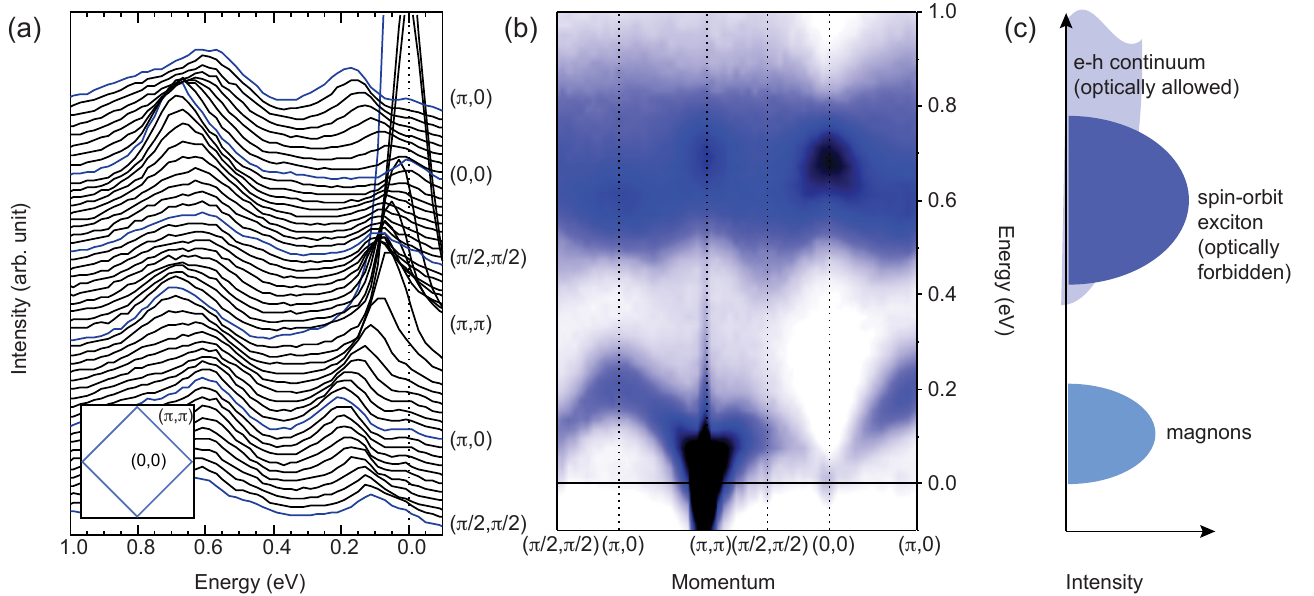}}
\vspace*{-0.3cm}%
%
\caption{(a) Energy loss spectra recorded at T=15 K, well below the
  T$_\textrm{N}$ $\approx$240 K~\cite{cao98,bjkim09}, along a path in the
  constant L=34 plane. The path was chosen to avoid the magnetic Bragg peaks,
  which appear at two of the four corners of the unfolded unit cell (black
  square) shown in the inset (where the same conventions as in Fig.~2 are
  used). 
(b) Image plot of the data
  shown in (a). (c) Schematic of the three representative features in the
  data.}\label{fig:fig3} 
\end{figure*}

\SIO has a canted antiferromagnetic (AF) structure~\cite{bjkim09} with T$_{\textrm N}\approx240$ K~\cite{cao98}, as shown in Fig. 1(b). Although the `internal' structure of a single magnetic moment in \SIOns, composed of orbital and spin, is drastically different from that of pure spins in \LCOns, a parent insulator for cuprate superconductors, the two compounds share apparently similar magnetic structure. Figures~\ref{fig:fig2}(a) and (b) show the dispersion and intensity, respectively, of the single magnon extracted by fitting energy distribution curves shown in Fig.~\ref{fig:fig3}(a). We highlight three important observations. First, not only the dispersion but also the momentum dependence of the intensity show striking similarities to those observed in the cuprates by inelastic neutron scattering, for instance in \LCOns~\cite{coldea01}. This provides confidence that the observed mode is indeed a single magnon excitation~\cite{amentprb11,ament09,haverkort10,braicovich10}. Using hard x-ray RIXS allows mapping of an entire Brillouin zone within only a few degrees of 90$
^\circ$ scattering geometry so that the spectrum reveals the intrinsic dynamical structural factor with minimal RIXS matrix element effects. Second, the measured magnon dispersion relation strongly supports the theories predicting that the superexhange interactions of isospins on a square lattice with corner-sharing octahedra are governed by a SU(2) invariant Hamiltonian with AF coupling~\cite{jackeli09,wang11}. Third, the magnon mode in \SIO has a bandwidth of $\sim$200~meV as compared to $\sim$300~meV in \LCO~\cite{coldea01} and Sr$_2$CuO$_2$Cl$_2$~\cite{guarise10}, which is consistent with energy scales of hopping $t$ and on-site Coulomb energy $U$ in \SIO being smaller by roughly 50$\%$ than those reported for the cuprates~\cite{wang11,jin09,watanabe10}.

%
%

For a quantitative description, we have fitted the magnon dispersion using a phenomenological $J$-$J'$-$J''$ model~\cite{comm2}. Here, the $J$, $J'$, and $J''$ correspond to the first, second, and third nearest neighbors, respectively. In this model, the downward dispersion along the magnetic Brillouin zone from ($\pi$,0) to ($\pi$/2, $\pi$/2) is accounted for by a ferromagnetic $J'$~\cite{coldea01,comm2}. We find $J$=60, $J'$=-20, and $J''$=15~meV. The nearest neighbor interaction $J$ is again smaller than found in cuprates by roughly 50$\%$. The fit can be improved by including higher order terms from longer-range interactions, which were also found to be important in the case of Sr$_2$CuO$_2$Cl$_2$~\cite{guarise10}. However, here we do not pursue this path because, as we show below, another kind of magnetic modes in \SIOns, which is not present in cuprates, may affect the magnon dispersion.

%
%

\begin{figure}[t]
\vspace*{0.1cm}\centerline{\includegraphics[width=0.95\columnwidth,angle=0]{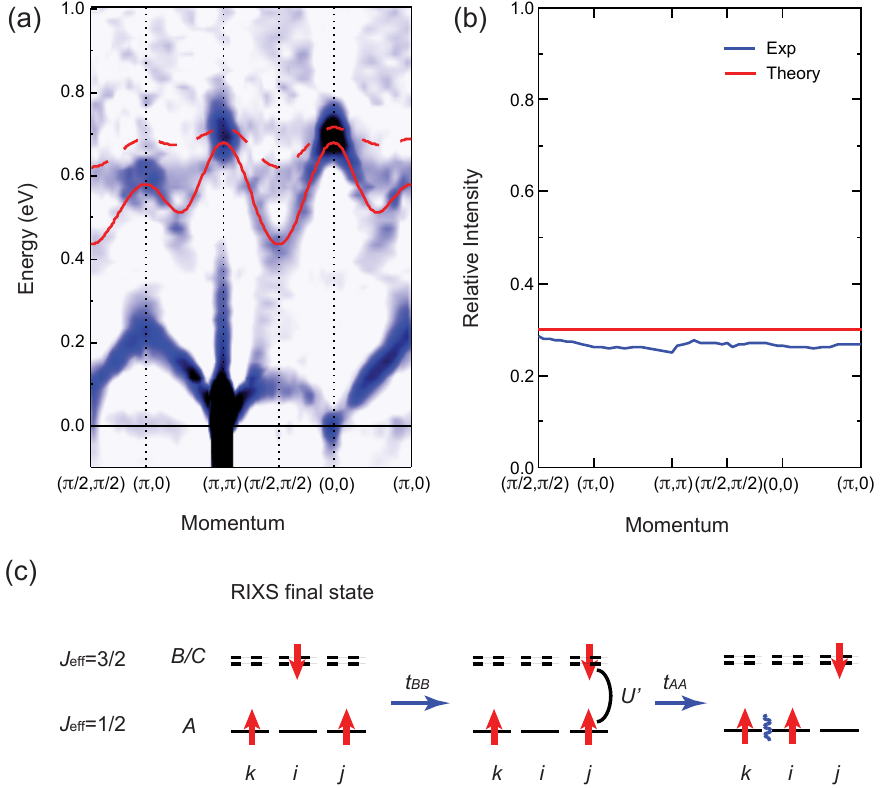}}
\vspace*{-0.5cm}%
%
\caption{(a) Second derivative of the data shown in Fig.~\ref{fig:fig3}(b), overlaid with the calculation of the dispersion (red solid and dashed lines). The upper branch (dashed line) has less spectral weight than the lower branch (solid line) ~\cite{supp}. W=63 meV was chosen in the calculation, which is within the estimated range ~\cite{supp}. (b) Comparison of the experimental and theoretical integrated spectral weight for the two spin-orbit exciton modes normalized to the single magnon mode at ($\pi$,0). (c) Schematic of the spin-orbit exciton hopping in the hole representation. By the RIXS process, the hole in the $i$ site is excited to the B/C doublets, which are the \jeff=3/2 quartet split by the tetragonal crystal field. This excited hole hops to the neighboring site $j$ with the intermediate energy of $U’$. The other hole in the $j$ site hops back to the $i$ site, thereby completing the spin-orbit exciton hopping processes and also creating a magnon (blue wavy line).}\label{fig:fig4}
\end{figure}

Ultimately, characterizing the magnon mode is important because it strongly
renormalizes the dispersion of a doped hole/electron, and is believed to
provide a pairing mechanism for HSTC. We now show that \SIO\ supports an
exciton mode, which gives access to the dynamics of a particle propagating in
the background of AF ordered moments even in an undoped case. Figures~\ref{fig:fig3}(a) and (b) show the energy loss spectra along high symmetry directions. No corrections to the raw data such as normalization or subtraction of the elastic contaminations have been made. Another virtue of using hard x-ray is that by working in the vicinity of 90$^\circ$ scattering geomeotry elastic (Thompson) scattering can be strongly suppressed. In addition to low-energy magnon branch ($\leq$ 0.2 eV), we observe high-energy excitations with strong momentum dependence in the energy range of 0.4$\sim$0.8~eV. This mode is superimposed on top of a continuum generated by particle-hole excitations across the Mott gap (estimated to be $\approx$0.4 eV from optical spectroscopy~\cite{moon09}). This is schematically shown in Fig.~\ref{fig:fig3}(c). Taking the second derivative of the raw data de-emphasizes the intensity arising from the particle-hole continuum and reveals a clear dispersive feature above 0.4 eV, as shown in Fig.~\ref{fig:fig4}(a). The energy scale of this excitation coincides with the known energy scale of spin-orbit coupling in \SIO($\zeta_{SO}\sim$0.5 eV)~\cite{bjkim08}, and thus we assign it to intra-site excitations of a hole across the spin-orbit split levels in the $t_{2g}$ manifold, i.e. from the \jeff=1/2 level to the one of the \jeff=3/2 levels~\cite{bjkim08,amentprb11,supp} (see Fig.~\ref{fig:fig4}(c)). We refer to such an excitation as `spin-orbit exciton'~\cite{holden71}.

The dispersion of spin-orbit exciton with a bandwidth of at least 0.3 eV implies that this local excitation can propagate coherently through the lattice. Our model of the spin-orbit exciton starts from a recognition that the hopping process is formally analogous to the problem of a hole propagating in the background of AF ordered moments, which has been extensively studied in the context of cuprate HTSC~\cite{leeRMP06}. Although the spin-orbit exciton does not carry a charge, its hopping creates a trail of “misaligned” spins and thus is subject to the same kind of renormalization by magnons as that experienced by a doped hole~\cite{schmitt88}. It is well known that the dispersion of a doped hole in cuprates has a minimum at ($\pi$/2,$\pi$/2)~\cite{wells98}, i.e., at the AF magnetic Brillouin zone boundary. Since \SIO\ has a similar magnetic order~\cite{bjkim09}, it can be understood by analogy that the dispersion of the spin-orbit exciton should also have its minimum at ($\pi$/2,$\pi$/2).

The overall bandwidth is determined by the parameters involved in the hopping process, which is depicted in Fig.~\ref{fig:fig4}(c) in the hole picture. It involves moving an excited hole to a neighboring site, which happens in two steps. First, the excited hole in site $i$ hops to a neighboring site $j$, generating an intermediate state with energy $U'$, which is the Coulomb repulsion between two holes at a site in two different spin-orbital quantum levels. Then, the other hole in site $j$ hops back to site $i$. Thus, the energy scale of the dispersion is set by $2t_{AA}t_{BB}/U'$, which is of the order of the magnetic exchange couplings. In fact, these processes lead to the superexchange interactions responsible for the magnetic ordering, but here they involve both the ground state and excited states of Ir ions.

Technically, the spin-orbit exciton hopping can be described by the following Hamiltonian
\begin{equation}
H=-\sum_{i,j} W^{\alpha\beta}_{i,j} X_{i\alpha}^\dagger X_{j\beta} (b_j^\dagger+b_i),
\end{equation}
where $i$ indicates the lattice site, $b$ ($b^\dagger$) is the magnon annihilation (creation) operator, and $X$ denotes the spin-orbit exciton that carries a quantum number $\alpha$ belonging either to the $B$ or $C$ doublet in the \jeff=3/2 manifold (see Fig.~\ref{fig:fig3}(c)). From this expression, the analogy with the case of a moving hole is apparent; in place of the hopping $t$ for the doped hole, we have an effective spin-orbit hopping matrix $W^{αβ}$ with its overall energy
scale set by $W = 2t^2/U’$. 

We calculated the spin-orbit exciton dispersion by evaluating its self-energy matrix
expressed as 
\begin{equation}
\Sigma^{\alpha\beta}_{\mathbf k}=-z^2W^2
\sum_{\gamma, \mathbf q} \frac{M^{\alpha\gamma}_{\mathbf k,q} 
M^{\gamma\beta}_{\mathbf k,q}}{\omega_{\mathbf q}},  
\end{equation}
where $z$ is the coordination number and $M$ denotes the vertex ~\cite{supp}, using the actual experimental magnon dispersion relation for $\omega_{\mathbf q}$ as shown in Fig.~\ref{fig:fig2}(a). The only adjustable parameter is $W$, which only contributes to the overall scaling of the dispersion. The model correctly captures the main features of the data: the locations of extrema in the dispersion (Fig.~\ref{fig:fig4}(a)), nearly momentum independent integrated spectral weight (Fig.~\ref{fig:fig4}(b)), and the intensity relative to the magnon intensity (Fig.~\ref{fig:fig4}(b)). The theory described above is developed in {\it Supplementary Material}.

%
%

Our measurement of the spin-orbit exciton dispersion has important
implications in modeling 5$d$ transition-metal oxides with strong SOC. First,
it shows that not only the \jeff=1/2 states are localized but also the
\jeff=3/2 states largely retain their atomic-like character. In a contrasting
model, in which \jeff=3/2 states form an itinerant band and only the \jeff=1/2
states are localized, much akin to the orbital-selective Mott transition
scenario~\cite{anisimov02}, one expects to see only an electron-hole continuum
resulting from the independent propagations of a hole and an
electron. Instead, we see the spin-orbit exciton coexisting with the
particle-hole continuum; a duality of atomic and band nature of the same 5$d$
electrons. Second, the existence of the virtually bound \jeff=3/2 states only
$\sim$0.5 eV above the ground state implies that the superexchange
interactions entail multiorbital contributions. Thus, even for an apparently
single orbital \jeff=1/2 system such as Sr$_2$IrO$_4$, the magnetic interactions are
multiorbital in character, a fact that must to be taken into account in any
quantitative model. 

Despite such important differences in the high energy scale, our measurement of the magnon spectrum highlights the similarities with cuprates in the low energy effective physics - a rare realization of (iso)spin one-half moments on a square lattice with Heisenberg SU(2) invariant interactions and comparable magnon bandwidth. Further, from the observed spin-orbit exciton dispersion, we may expect that a doped hole/electron in \SIO will display the same dynamics as that observed for a doped hole/electron in the cuprates. The phase diagram of lightly doped \SIO has just begun to be revealed experimentally~\cite{korneta10,gepreprint10}. Although superconductivity has not yet been reported, some anomalies that bear strong resemblance to cuprates such as T-linear resistivity have been seen~\cite{gepreprint10,OkabePRB11}. Only further study will tell if doping can drive \SIO superconducting.

\acknowledgements{B. J. K. thanks T. Senthil and H. M. Ronnow for discussions. Work in the Material Science Division and the use of the Advanced Photon Source at the Argonne National Laboratory was supported by the U.S. DOE under Contract No. DE-AC02-06CH11357. Y. K. was supported by the Canada Foundation for Innovation, Ontario Research Fund, and Natural Sciences and Engineering Research Council of Canada. M. vV. was supported by the US Department of Energy (DOE), Office of Basic Energy Sciences, Division of Materials Sciences and Engineering under Award DE-FG02-03ER46097. This work benefited from the RIXS collaboration supported by the Computational Materials Science Network (CMSN) program of the Division of Materials Science and Engineering, Office of Basic Energy Sciences (BES), US DOE under grant number DE-FG02-08ER46540.}

\bibliography{SIO214REF_JHK}

\begin{thebibliography}{10}%
\makeatletter
\providecommand \@ifxundefined [1]{%
 \ifx #1\undefined \expandafter \@firstoftwo
 \else \expandafter \@secondoftwo
\fi
}%
\providecommand \@ifnum [1]{%
 \ifnum #1\expandafter \@firstoftwo
 \else \expandafter \@secondoftwo
\fi
}%
\providecommand \enquote [1]{``#1''}%
\providecommand \bibnamefont  [1]{#1}%
\providecommand \bibfnamefont [1]{#1}%
\providecommand \citenamefont [1]{#1}%
\providecommand\href[0]{\@sanitize\@href}%
\providecommand\@href[1]{\endgroup\@@startlink{#1}\endgroup\@@href}%
\providecommand\@@href[1]{#1\@@endlink}%
\providecommand \@sanitize [0]{\begingroup\catcode`\&12\catcode`\#12\relax}%
\@ifxundefined \pdfoutput {\@firstoftwo}{%
 \@ifnum{\z@=\pdfoutput}{\@firstoftwo}{\@secondoftwo}%
}{%
 \providecommand\@@startlink[1]{\leavevmode\special{html:<a href="#1">}}%
 \providecommand\@@endlink[0]{\special{html:</a>}}%
}{%
 \providecommand\@@startlink[1]{%
  \leavevmode
  \pdfstartlink
   attr{/Border[0 0 1 ]/H/I/C[0 1 1]}%
   user{/Subtype/Link/A<</Type/Action/S/URI/URI(#1)>>}%
  \relax
 }%
 \providecommand\@@endlink[0]{\pdfendlink}%
}%
\providecommand \url  [0]{\begingroup\@sanitize \@url }%
\providecommand \@url [1]{\endgroup\@href {#1}{\urlprefix}}%
\providecommand \urlprefix [0]{URL }%
\providecommand \Eprint[0]{\href }%
\@ifxundefined \urlstyle {%
  \providecommand \doi [1]{doi:\discretionary{}{}{}#1}%
}{%
  \providecommand \doi [0]{doi:\discretionary{}{}{}\begingroup
  \urlstyle{rm}\Url }%
}%
\providecommand \doibase [0]{http://dx.doi.org/}%
\providecommand \Doi[1]{\href{\doibase#1}}%
\providecommand \bibAnnote [3]{%
  \BibitemShut{#1}%
  \begin{quotation}\noindent
    \textsc{Key:}\ #2\\\textsc{Annotation:}\ #3%
  \end{quotation}%
}%
\providecommand \bibAnnoteFile [2]{%
  \IfFileExists{#2}{\bibAnnote {#1} {#2} {\input{#2}}}{}%
}%
\providecommand \typeout [0]{\immediate \write \m@ne }%
\providecommand \selectlanguage [0]{\@gobble}%
\providecommand \bibinfo [0]{\@secondoftwo}%
\providecommand \bibfield [0]{\@secondoftwo}%
\providecommand \translation [1]{[#1]}%
\providecommand \BibitemOpen[0]{}%
\providecommand \bibitemStop [0]{}%
\providecommand \bibitemNoStop [0]{.\EOS\space}%
\providecommand \EOS [0]{\spacefactor3000\relax}%
\providecommand \BibitemShut [1]{\csname bibitem#1\endcsname}%
\bibitem{GoodEnough}%
  \BibitemOpen
  \bibfield{author}{%
  \bibinfo {author} {\bibfnamefont{J.~B.}\ \bibnamefont{Goodenough}},\ }%
  \emph{\bibinfo {title} {Magnetism and the Chemical Bond}}\ (\bibinfo
  {publisher} {Interscience, New York},\ \bibinfo {year} {1963})\ p.\ \bibinfo
  {pages} {1329}%
  \bibAnnoteFile{NoStop}{GoodEnough}%
\bibitem{jackeli09}%
  \BibitemOpen
  \bibfield{author}{%
  \bibinfo {author} {\bibfnamefont{G.}~\bibnamefont{Jackeli}}\ and\ \bibinfo
  {author} {\bibfnamefont{G.}~\bibnamefont{Khaliullin}},\ }%
  \bibfield{journal}{%
  \bibinfo {journal} {Phys. Rev. Lett.}\ }%
  \textbf{\bibinfo {volume} {102}},\ \bibinfo {pages} {017205} (\bibinfo {year}
  {2009})%
  \bibAnnoteFile{NoStop}{jackeli09}%
\bibitem{chaloupka10}%
  \BibitemOpen
  \bibfield{author}{%
  \bibinfo {author} {\bibfnamefont{J.}~\bibnamefont{Chaloupka}}, \bibinfo
  {author} {\bibfnamefont{G.}~\bibnamefont{Jackeli}},\ and\ \bibinfo {author}
  {\bibfnamefont{G.}~\bibnamefont{Khaliullin}},\ }%
  \bibfield{journal}{%
  \bibinfo {journal} {Phys. Rev. Lett.}\ }%
  \textbf{\bibinfo {volume} {105}},\ \bibinfo {pages} {027204} (\bibinfo {year}
  {2010})%
  \bibAnnoteFile{NoStop}{chaloupka10}%
\bibitem{HillPRB11}%
  \BibitemOpen
  \bibfield{author}{%
  \bibinfo {author} {\bibfnamefont{X.}~\bibnamefont{Liu}}, \bibinfo {author}
  {\bibfnamefont{T.}~\bibnamefont{Berlijn}}, \bibinfo {author}
  {\bibfnamefont{W.-G.}\ \bibnamefont{Yin}}, \bibinfo {author}
  {\bibfnamefont{W.}~\bibnamefont{Ku}}, \bibinfo {author}
  {\bibfnamefont{A.}~\bibnamefont{Tsvelik}}, \bibinfo {author}
  {\bibfnamefont{Y.-J.}\ \bibnamefont{Kim}}, \bibinfo {author}
  {\bibfnamefont{H.}~\bibnamefont{Gretarssona}}, \bibinfo {author}
  {\bibfnamefont{Y.}~\bibnamefont{Singh}}, \bibinfo {author}
  {\bibfnamefont{P.}~\bibnamefont{Gegenwart}},\ and\ \bibinfo {author}
  {\bibfnamefont{J.~P.}\ \bibnamefont{Hill}},\ }%
  \bibfield{journal}{%
  \bibinfo {journal} {Phys. Rev. B}\ }%
  \textbf{\bibinfo {volume} {83}},\ \bibinfo {pages} {220403} (\bibinfo {year}
  {2011})%
  \bibAnnoteFile{NoStop}{HillPRB11}%
\bibitem{BalentsNPhys10}%
  \BibitemOpen
  \bibfield{author}{%
  \bibinfo {author} {\bibfnamefont{D.}~\bibnamefont{Pesin}}\ and\ \bibinfo
  {author} {\bibfnamefont{L.}~\bibnamefont{Balents}},\ }%
  \bibfield{journal}{%
  \bibinfo {journal} {Nat. Phys.}\ }%
  \textbf{\bibinfo {volume} {6}},\ \bibinfo {pages} {376} (\bibinfo {year}
  {2010})%
  \bibAnnoteFile{NoStop}{BalentsNPhys10}%
\bibitem{WanPRB11}%
  \BibitemOpen
  \bibfield{author}{%
  \bibinfo {author} {\bibfnamefont{X.}~\bibnamefont{Wan}}, \bibinfo {author}
  {\bibfnamefont{A.}~\bibnamefont{Turner}}, \bibinfo {author}
  {\bibfnamefont{A.}~\bibnamefont{Vishwanath}},\ and\ \bibinfo {author}
  {\bibfnamefont{S.~Y.}\ \bibnamefont{Savrasov}},\ }%
  \bibfield{journal}{%
  \bibinfo {journal} {Phys. Rev. B}\ }%
  \textbf{\bibinfo {volume} {83}},\ \bibinfo {pages} {205101} (\bibinfo {year}
  {2011})%
  \bibAnnoteFile{NoStop}{WanPRB11}%
\bibitem{bjkim08}%
  \BibitemOpen
  \bibfield{author}{%
  \bibinfo {author} {\bibfnamefont{B.~J.}\ \bibnamefont{Kim}}, \bibinfo
  {author} {\bibfnamefont{H.}~\bibnamefont{Jin}}, \bibinfo {author}
  {\bibfnamefont{S.~J.}\ \bibnamefont{Moon}}, \bibinfo {author}
  {\bibfnamefont{J.-Y.}\ \bibnamefont{Kim}}, \bibinfo {author}
  {\bibfnamefont{B.-G.}\ \bibnamefont{Park}}, \bibinfo {author}
  {\bibfnamefont{C.~S.}\ \bibnamefont{Leem}}, \bibinfo {author}
  {\bibfnamefont{J.}~\bibnamefont{Yu}}, \bibinfo {author}
  {\bibfnamefont{T.~W.}\ \bibnamefont{Noh}}, \bibinfo {author}
  {\bibfnamefont{C.}~\bibnamefont{Kim}}, \bibinfo {author}
  {\bibfnamefont{S.-J.}\ \bibnamefont{Oh}}, \bibinfo {author}
  {\bibfnamefont{J.-H.}\ \bibnamefont{Park}}, \bibinfo {author}
  {\bibfnamefont{V.}~\bibnamefont{Durairaj}}, \bibinfo {author}
  {\bibfnamefont{G.}~\bibnamefont{Cao}},\ and\ \bibinfo {author}
  {\bibfnamefont{E.}~\bibnamefont{Rotenberg}},\ }%
  \bibfield{journal}{%
  \bibinfo {journal} {Phys. Rev. Lett.}\ }%
  \textbf{\bibinfo {volume} {101}},\ \bibinfo {pages} {076402} (\bibinfo {year}
  {2008})%
  \bibAnnoteFile{NoStop}{bjkim08}%
\bibitem{bjkim09}%
  \BibitemOpen
  \bibfield{author}{%
  \bibinfo {author} {\bibfnamefont{B.~J.}\ \bibnamefont{Kim}}, \bibinfo
  {author} {\bibfnamefont{H.}~\bibnamefont{Ohsumi}}, \bibinfo {author}
  {\bibfnamefont{T.}~\bibnamefont{Komesu}}, \bibinfo {author}
  {\bibfnamefont{S.}~\bibnamefont{Sakai}}, \bibinfo {author}
  {\bibfnamefont{T.}~\bibnamefont{Morita}}, \bibinfo {author}
  {\bibfnamefont{H.}~\bibnamefont{Takagi}},\ and\ \bibinfo {author}
  {\bibfnamefont{T.}~\bibnamefont{Arima}},\ }%
  \bibfield{journal}{%
  \bibinfo {journal} {Science}\ }%
  \textbf{\bibinfo {volume} {323}},\ \bibinfo {pages} {1329} (\bibinfo {year}
  {2009})%
  \bibAnnoteFile{NoStop}{bjkim09}%
\bibitem{crawford94}%
  \BibitemOpen
  \bibfield{author}{%
  \bibinfo {author} {\bibfnamefont{M.~K.}\ \bibnamefont{Crawford}}, \bibinfo
  {author} {\bibfnamefont{M.~A.}\ \bibnamefont{Subramanian}}, \bibinfo {author}
  {\bibfnamefont{R.~L.}\ \bibnamefont{Harlow}}, \bibinfo {author}
  {\bibfnamefont{J.~A.}\ \bibnamefont{Fernandez-Baca}}, \bibinfo {author}
  {\bibfnamefont{Z.~R.}\ \bibnamefont{Wang}},\ and\ \bibinfo {author}
  {\bibfnamefont{D.~C.}\ \bibnamefont{Johnston}},\ }%
  \bibfield{journal}{%
  \bibinfo {journal} {Phys. Rev. B}\ }%
  \textbf{\bibinfo {volume} {49}},\ \bibinfo {pages} {9198} (\bibinfo {year}
  {1994})%
  \bibAnnoteFile{NoStop}{crawford94}%
\bibitem{wang11}%
  \BibitemOpen
  \bibfield{author}{%
  \bibinfo {author} {\bibfnamefont{F.}~\bibnamefont{Wang}}\ and\ \bibinfo
  {author} {\bibfnamefont{T.}~\bibnamefont{Senthil}},\ }%
  \bibfield{journal}{%
  \bibinfo {journal} {Phys. Rev. Lett.}\ }%
  \textbf{\bibinfo {volume} {106}},\ \bibinfo {pages} {136402} (\bibinfo {year}
  {2011})%
  \bibAnnoteFile{NoStop}{wang11}%
\bibitem{ament11}%
  \BibitemOpen
  \bibfield{author}{%
  \bibinfo {author} {\bibfnamefont{L.~J.~P.}\ \bibnamefont{Ament}}, \bibinfo
  {author} {\bibfnamefont{M.}~\bibnamefont{van Veenendaal}}, \bibinfo {author}
  {\bibfnamefont{T.~P.}\ \bibnamefont{Devereaux}}, \bibinfo {author}
  {\bibfnamefont{J.~P.}\ \bibnamefont{Hill}},\ and\ \bibinfo {author}
  {\bibfnamefont{J.}~\bibnamefont{van~den Brink}},\ }%
  \bibfield{journal}{%
  \bibinfo {journal} {Rev. Mod. Phys.}\ }%
  \textbf{\bibinfo {volume} {83}},\ \bibinfo {pages} {705} (\bibinfo {year}
  {2011})%
  \bibAnnoteFile{NoStop}{ament11}%
\bibitem{cao98}%
  \BibitemOpen
  \bibfield{author}{%
  \bibinfo {author} {\bibfnamefont{G.}~\bibnamefont{Cao}}, \bibinfo {author}
  {\bibfnamefont{J.}~\bibnamefont{Bolivar}}, \bibinfo {author}
  {\bibfnamefont{S.}~\bibnamefont{McCall}}, \bibinfo {author}
  {\bibfnamefont{J.~E.}\ \bibnamefont{Crow}},\ and\ \bibinfo {author}
  {\bibfnamefont{R.~P.}\ \bibnamefont{Guertin}},\ }%
  \bibfield{journal}{%
  \bibinfo {journal} {Phys. Rev. B}\ }%
  \textbf{\bibinfo {volume} {57}},\ \bibinfo {pages} {R11039} (\bibinfo {year}
  {1998})%
  \bibAnnoteFile{NoStop}{cao98}%
\bibitem{coldea01}%
  \BibitemOpen
  \bibfield{author}{%
  \bibinfo {author} {\bibfnamefont{R.}~\bibnamefont{Coldea}}, \bibinfo {author}
  {\bibfnamefont{S.~M.}\ \bibnamefont{Hayden}}, \bibinfo {author}
  {\bibfnamefont{G.}~\bibnamefont{Aeppli}}, \bibinfo {author}
  {\bibfnamefont{T.~G.}\ \bibnamefont{Perring}}, \bibinfo {author}
  {\bibfnamefont{C.~D.}\ \bibnamefont{Frost}}, \bibinfo {author}
  {\bibfnamefont{T.~E.}\ \bibnamefont{Mason}}, \bibinfo {author}
  {\bibfnamefont{S.-W.}\ \bibnamefont{Cheong}},\ and\ \bibinfo {author}
  {\bibfnamefont{Z.}~\bibnamefont{Fisk}},\ }%
  \bibfield{journal}{%
  \bibinfo {journal} {Phys. Rev. Lett.}\ }%
  \textbf{\bibinfo {volume} {86}},\ \bibinfo {pages} {5377} (\bibinfo {year}
  {2001})%
  \bibAnnoteFile{NoStop}{coldea01}%
\bibitem{amentprb11}%
  \BibitemOpen
  \bibfield{author}{%
  \bibinfo {author} {\bibfnamefont{L.~J.~P.}\ \bibnamefont{Ament}}, \bibinfo
  {author} {\bibfnamefont{G.}~\bibnamefont{Khaliullin}},\ and\ \bibinfo
  {author} {\bibfnamefont{J.}~\bibnamefont{van~den Brink}},\ }%
  \bibfield{journal}{%
  \bibinfo {journal} {Phys. Rev. B}\ }%
  \textbf{\bibinfo {volume} {84}},\ \bibinfo {pages} {020403} (\bibinfo {year}
  {2011})%
  \bibAnnoteFile{NoStop}{amentprb11}%
\bibitem{ament09}%
  \BibitemOpen
  \bibfield{author}{%
  \bibinfo {author} {\bibfnamefont{L.~J.~P.}\ \bibnamefont{Ament}}, \bibinfo
  {author} {\bibfnamefont{G.}~\bibnamefont{Ghiringhelli}}, \bibinfo {author}
  {\bibfnamefont{M.~M.}\ \bibnamefont{Sala}}, \bibinfo {author}
  {\bibfnamefont{L.}~\bibnamefont{Braicovich}},\ and\ \bibinfo {author}
  {\bibfnamefont{J.}~\bibnamefont{van~den Brink}},\ }%
  \bibfield{journal}{%
  \bibinfo {journal} {Phys. Rev. Lett.}\ }%
  \textbf{\bibinfo {volume} {103}},\ \bibinfo {pages} {117003} (\bibinfo {year}
  {2009})%
  \bibAnnoteFile{NoStop}{ament09}%
\bibitem{haverkort10}%
  \BibitemOpen
  \bibfield{author}{%
  \bibinfo {author} {\bibfnamefont{M.~W.}\ \bibnamefont{Haverkort}},\ }%
  \bibfield{journal}{%
  \bibinfo {journal} {Phys. Rev. Lett.}\ }%
  \textbf{\bibinfo {volume} {105}},\ \bibinfo {pages} {167404} (\bibinfo {year}
  {2010})%
  \bibAnnoteFile{NoStop}{haverkort10}%
\bibitem{braicovich10}%
  \BibitemOpen
  \bibfield{author}{%
  \bibinfo {author} {\bibfnamefont{L.}~\bibnamefont{Braicovich}}, \bibinfo
  {author} {\bibfnamefont{J.}~\bibnamefont{van~den Brink}}, \bibinfo {author}
  {\bibfnamefont{V.}~\bibnamefont{Bisogni}}, \bibinfo {author}
  {\bibfnamefont{M.~M.}\ \bibnamefont{Sala}}, \bibinfo {author}
  {\bibfnamefont{L.~J.~P.}\ \bibnamefont{Ament}}, \bibinfo {author}
  {\bibfnamefont{N.~B.}\ \bibnamefont{Brookes}}, \bibinfo {author}
  {\bibfnamefont{G.~M.~D.}\ \bibnamefont{Luca}}, \bibinfo {author}
  {\bibfnamefont{M.}~\bibnamefont{Salluzzo}}, \bibinfo {author}
  {\bibfnamefont{T.}~\bibnamefont{Schmitt}}, \bibinfo {author}
  {\bibfnamefont{V.~N.}\ \bibnamefont{Strocov}},\ and\ \bibinfo {author}
  {\bibfnamefont{G.}~\bibnamefont{Ghiringhelli}},\ }%
  \bibfield{journal}{%
  \bibinfo {journal} {Phys. Rev. Lett.}\ }%
  \textbf{\bibinfo {volume} {104}},\ \bibinfo {pages} {077002} (\bibinfo {year}
  {2010})%
  \bibAnnoteFile{NoStop}{braicovich10}%
\bibitem{guarise10}%
  \BibitemOpen
  \bibfield{author}{%
  \bibinfo {author} {\bibfnamefont{M.}~\bibnamefont{Guarise}}, \bibinfo
  {author} {\bibfnamefont{B.~D.}\ \bibnamefont{Piazza}}, \bibinfo {author}
  {\bibfnamefont{M.~M.}\ \bibnamefont{Sala}}, \bibinfo {author}
  {\bibfnamefont{G.}~\bibnamefont{Ghiringhelli}}, \bibinfo {author}
  {\bibfnamefont{L.}~\bibnamefont{Braicovich}}, \bibinfo {author}
  {\bibfnamefont{H.}~\bibnamefont{Berger}}, \bibinfo {author}
  {\bibfnamefont{J.~N.}\ \bibnamefont{Hancock}}, \bibinfo {author}
  {\bibfnamefont{D.}~\bibnamefont{van~der Marel}}, \bibinfo {author}
  {\bibfnamefont{T.}~\bibnamefont{Schmitt}}, \bibinfo {author}
  {\bibfnamefont{V.~N.}\ \bibnamefont{Strocov}}, \bibinfo {author}
  {\bibfnamefont{L.~J.~P.}\ \bibnamefont{Ament}}, \bibinfo {author}
  {\bibfnamefont{J.}~\bibnamefont{van~den Brink}}, \bibinfo {author}
  {\bibfnamefont{P.-H.}\ \bibnamefont{Lin}}, \bibinfo {author}
  {\bibfnamefont{P.}~\bibnamefont{Xu}}, \bibinfo {author}
  {\bibfnamefont{H.~M.}\ \bibnamefont{R{\o}nnow}},\ and\ \bibinfo {author}
  {\bibfnamefont{M.}~\bibnamefont{Grioni}},\ }%
  \bibfield{journal}{%
  \bibinfo {journal} {Phys. Rev. Lett.}\ }%
  \textbf{\bibinfo {volume} {105}},\ \bibinfo {pages} {157006} (\bibinfo {year}
  {2010})%
  \bibAnnoteFile{NoStop}{guarise10}%
\bibitem{jin09}%
  \BibitemOpen
  \bibfield{author}{%
  \bibinfo {author} {\bibfnamefont{H.}~\bibnamefont{Jin}}, \bibinfo {author}
  {\bibfnamefont{H.}~\bibnamefont{Jeong}}, \bibinfo {author}
  {\bibfnamefont{T.}~\bibnamefont{Ozaki}},\ and\ \bibinfo {author}
  {\bibfnamefont{J.}~\bibnamefont{Yu}},\ }%
  \bibfield{journal}{%
  \bibinfo {journal} {Phys. Rev. B}\ }%
  \textbf{\bibinfo {volume} {80}},\ \bibinfo {pages} {075112} (\bibinfo {year}
  {2009})%
  \bibAnnoteFile{NoStop}{jin09}%
\bibitem{watanabe10}%
  \BibitemOpen
  \bibfield{author}{%
  \bibinfo {author} {\bibfnamefont{H.}~\bibnamefont{Watanabe}}, \bibinfo
  {author} {\bibfnamefont{T.}~\bibnamefont{Shirakawa}},\ and\ \bibinfo {author}
  {\bibfnamefont{S.}~\bibnamefont{Yunoki}},\ }%
  \bibfield{journal}{%
  \bibinfo {journal} {Phys. Rev. Lett.}\ }%
  \textbf{\bibinfo {volume} {105}},\ \bibinfo {pages} {216410} (\bibinfo {year}
  {2010})%
  \bibAnnoteFile{NoStop}{watanabe10}%
\bibitem{supp}%
  \BibitemOpen
  \bibinfo {note} {See supplementary material for details.}%
  \bibAnnoteFile{Stop}{supp}%
\bibitem{comm2}%
  \BibitemOpen
  \bibinfo {note} {We do not include the cyclic exchange J$_C$ because magnons
  cannot distinguish between ferromagnetic $J'$ and $J_C$. See, for example,
  Ref.~\onlinecite{toader05}}%
  \bibAnnoteFile{NoStop}{comm2}%
\bibitem{moon09}%
  \BibitemOpen
  \bibfield{author}{%
  \bibinfo {author} {\bibfnamefont{S.~J.}\ \bibnamefont{Moon}}, \bibinfo
  {author} {\bibfnamefont{H.}~\bibnamefont{Jin}}, \bibinfo {author}
  {\bibfnamefont{W.~S.}\ \bibnamefont{Choi}}, \bibinfo {author}
  {\bibfnamefont{J.~S.}\ \bibnamefont{Lee}}, \bibinfo {author}
  {\bibfnamefont{S.~S.~A.}\ \bibnamefont{Seo}}, \bibinfo {author}
  {\bibfnamefont{J.}~\bibnamefont{Yu}}, \bibinfo {author}
  {\bibfnamefont{G.}~\bibnamefont{Cao}}, \bibinfo {author}
  {\bibfnamefont{T.~W.}\ \bibnamefont{Noh}},\ and\ \bibinfo {author}
  {\bibfnamefont{Y.~S.}\ \bibnamefont{Lee}},\ }%
  \bibfield{journal}{%
  \bibinfo {journal} {Phys. Rev. B}\ }%
  \textbf{\bibinfo {volume} {80}},\ \bibinfo {pages} {195110} (\bibinfo {year}
  {2009})%
  \bibAnnoteFile{NoStop}{moon09}%
\bibitem{holden71}%
  \BibitemOpen
  \bibfield{author}{%
  \bibinfo {author} {\bibfnamefont{T.~M.}\ \bibnamefont{Holden}}, \bibinfo
  {author} {\bibfnamefont{W.~J.~L.}\ \bibnamefont{Buyers}}, \bibinfo {author}
  {\bibfnamefont{E.~C.}\ \bibnamefont{Svensson}}, \bibinfo {author}
  {\bibfnamefont{R.~A.}\ \bibnamefont{Cowley}}, \bibinfo {author}
  {\bibfnamefont{M.~T.}\ \bibnamefont{Hutchings}}, \bibinfo {author}
  {\bibfnamefont{D.}~\bibnamefont{Hukin}},\ and\ \bibinfo {author}
  {\bibfnamefont{R.~W.~H.}\ \bibnamefont{Stevenson}},\ }%
  \bibfield{journal}{%
  \bibinfo {journal} {J. Phys. C}\ }%
  \textbf{\bibinfo {volume} {4}},\ \bibinfo {pages} {2127} (\bibinfo {year}
  {1971})%
  \bibAnnoteFile{NoStop}{holden71}%
\bibitem{leeRMP06}%
  \BibitemOpen
  \bibfield{author}{%
  \bibinfo {author} {\bibfnamefont{P.~A.}\ \bibnamefont{Lee}}, \bibinfo
  {author} {\bibfnamefont{N.}~\bibnamefont{Nagaosa}},\ and\ \bibinfo {author}
  {\bibfnamefont{X.-G.}\ \bibnamefont{Wen}},\ }%
  \bibfield{journal}{%
  \bibinfo {journal} {Rev. Mod. Phys.}\ }%
  \textbf{\bibinfo {volume} {78}},\ \bibinfo {pages} {17} (\bibinfo {year}
  {2006})%
  \bibAnnoteFile{NoStop}{leeRMP06}%
\bibitem{schmitt88}%
  \BibitemOpen
  \bibfield{author}{%
  \bibinfo {author} {\bibfnamefont{S.}~\bibnamefont{Schmitt-Rink}}, \bibinfo
  {author} {\bibfnamefont{C.~M.}\ \bibnamefont{Varma}},\ and\ \bibinfo {author}
  {\bibfnamefont{A.~E.}\ \bibnamefont{Ruckenstein}},\ }%
  \bibfield{journal}{%
  \bibinfo {journal} {Phys. Rev. Lett.}\ }%
  \textbf{\bibinfo {volume} {60}},\ \bibinfo {pages} {2793} (\bibinfo {year}
  {1988})%
  \bibAnnoteFile{NoStop}{schmitt88}%
\bibitem{wells98}%
  \BibitemOpen
  \bibfield{author}{%
  \bibinfo {author} {\bibfnamefont{B.~O.}\ \bibnamefont{Wells}}, \bibinfo
  {author} {\bibfnamefont{Z.~X.}\ \bibnamefont{Shen}}, \bibinfo {author}
  {\bibfnamefont{A.}~\bibnamefont{Matsuura}}, \bibinfo {author}
  {\bibfnamefont{D.~M.}\ \bibnamefont{King}}, \bibinfo {author}
  {\bibfnamefont{M.~A.}\ \bibnamefont{Kastner}}, \bibinfo {author}
  {\bibfnamefont{M.}~\bibnamefont{Greven}},\ and\ \bibinfo {author}
  {\bibfnamefont{R.~J.}\ \bibnamefont{Birgeneau}},\ }%
  \bibfield{journal}{%
  \bibinfo {journal} {Phys. Rev. Lett.}\ }%
  \textbf{\bibinfo {volume} {80}},\ \bibinfo {pages} {4245} (\bibinfo {year}
  {1998})%
  \bibAnnoteFile{NoStop}{wells98}%
\bibitem{anisimov02}%
  \BibitemOpen
  \bibfield{author}{%
  \bibinfo {author} {\bibfnamefont{V.~I.}\ \bibnamefont{Anisimov}}, \bibinfo
  {author} {\bibfnamefont{I.~A.}\ \bibnamefont{Nekrasov}}, \bibinfo {author}
  {\bibfnamefont{D.~E.}\ \bibnamefont{Kondakov}}, \bibinfo {author}
  {\bibfnamefont{T.~M.}\ \bibnamefont{Rice}},\ and\ \bibinfo {author}
  {\bibfnamefont{M.}~\bibnamefont{Sigrist}},\ }%
  \bibfield{journal}{%
  \bibinfo {journal} {Eur. Phys. J. B}\ }%
  \textbf{\bibinfo {volume} {25}},\ \bibinfo {pages} {191} (\bibinfo {year}
  {2002})%
  \bibAnnoteFile{NoStop}{anisimov02}%
\bibitem{korneta10}%
  \BibitemOpen
  \bibfield{author}{%
  \bibinfo {author} {\bibfnamefont{O.~B.}\ \bibnamefont{Korneta}}, \bibinfo
  {author} {\bibfnamefont{T.}~\bibnamefont{Qi}}, \bibinfo {author}
  {\bibfnamefont{S.}~\bibnamefont{Chikara}}, \bibinfo {author}
  {\bibfnamefont{S.}~\bibnamefont{Parkin}}, \bibinfo {author}
  {\bibfnamefont{L.~E.~D.}\ \bibnamefont{Long}}, \bibinfo {author}
  {\bibfnamefont{P.}~\bibnamefont{Schlottmann}},\ and\ \bibinfo {author}
  {\bibfnamefont{G.}~\bibnamefont{Cao}},\ }%
  \bibfield{journal}{%
  \bibinfo {journal} {Phys. Rev. B}\ }%
  \textbf{\bibinfo {volume} {82}},\ \bibinfo {pages} {115117} (\bibinfo {year}
  {2010})%
  \bibAnnoteFile{NoStop}{korneta10}%
\bibitem{gepreprint10}%
  \BibitemOpen
  \bibinfo {note} {M. Ge, T. F. Qi, O. B. Korneta, D. E. De Long, P.
  Schlottmann, W. P. Crummett, and G. Cao, arXiv:1106.2381v1}%
  \bibAnnoteFile{NoStop}{gepreprint10}%
\bibitem{OkabePRB11}%
  \BibitemOpen
  \bibfield{author}{%
  \bibinfo {author} {\bibfnamefont{H.}~\bibnamefont{Okabe}}, \bibinfo {author}
  {\bibfnamefont{N.}~\bibnamefont{Takeshita}}, \bibinfo {author}
  {\bibfnamefont{M.}~\bibnamefont{Isobe}}, \bibinfo {author}
  {\bibfnamefont{E.}~\bibnamefont{Takayama-Muromachi}}, \bibinfo {author}
  {\bibfnamefont{T.}~\bibnamefont{Muranaka}},\ and\ \bibinfo {author}
  {\bibfnamefont{J.}~\bibnamefont{Akimitsu}},\ }%
  \bibfield{journal}{%
  \bibinfo {journal} {Phys. Rev. B}\ }%
  \textbf{\bibinfo {volume} {84}},\ \bibinfo {pages} {115127} (\bibinfo {year}
  {2011})%
  \bibAnnoteFile{NoStop}{OkabePRB11}%
\bibitem{toader05}%
  \BibitemOpen
  \bibfield{author}{%
  \bibinfo {author} {\bibfnamefont{A.~M.}\ \bibnamefont{Toader}}, \bibinfo
  {author} {\bibfnamefont{J.~P.}\ \bibnamefont{Goff}}, \bibinfo {author}
  {\bibfnamefont{M.}~\bibnamefont{Roger}}, \bibinfo {author}
  {\bibfnamefont{N.}~\bibnamefont{Shannon}}, \bibinfo {author}
  {\bibfnamefont{J.~R.}\ \bibnamefont{Stewart}},\ and\ \bibinfo {author}
  {\bibfnamefont{M.}~\bibnamefont{Enderle}},\ }%
  \bibfield{journal}{%
  \bibinfo {journal} {Phys. Rev. Lett.}\ }%
  \textbf{\bibinfo {volume} {94}},\ \bibinfo {pages} {197202} (\bibinfo {year}
  {2005})%
  \bibAnnoteFile{NoStop}{toader05}%
\end{thebibliography}%

\end{document}


\title{Supplementary Material: \\Isospin Dynamics in Sr$_2$IrO$_4$: Forging Links to Cuprate Superconductivity}


\author{Jungho Kim, D. Casa, M. H. Upton, T. Gog, Young-June Kim, J. F. Mitchell, M. van Veenendaal, M. Daghofer, J. van den Brink, G. Khaliullin, B. J. Kim}




\maketitle


\section{RIXS process and effective operators}
We describe the RIXS process and derive the effective RIXS operators for the
particular case of Sr$_2$IrO$_4$$^{\textrm S1}$. The states we use are $A_{\pm}$
for $J_{\textrm{eff}}=1/2$ ground state doublet (where the sign takes the 
role of a isospin degree of freedom) and $B_{\pm}$ and $C_{\pm}$ Kramers 
doublets for $J_{\textrm{eff}}=3/2$ excited states (see Fig.~S1). The 
states $A_{\pm}$, $B_{\pm}$, and $C_{\pm}$ depend on spin orbit coupling 
$\lambda$ and tetragonal crystal field $\Delta$$^{\textrm S2}$. The energy 
splitting $\Delta$ between the $A_{\pm}$ states and the $B_{\pm}$/$C_{\pm}$ 
subsystem has a dominant contribution from spin-orbit coupling $\frac{3}{2}\lambda$  
but also may be contributed by coupling to the lattice; The 
charge density distribution in the excited $J_{\textrm{eff}}=3/2$ states is 
different from that in the ground state, and hence the orbital flip may cost
additional energy. In terms of electron counting, the $B$ and $C$ doublets 
are completely occupied, while the $A$ level is half filled. We now switch 
to the hole notation, in which $B$ and $C$ are empty. In the 
$A$ subsystem, superexchange interactions stabilize an alternating magnetic 
(isospin) order. We consider the low-energy Hilbert space with   
singly occupied sites with an empty $B$/$C$ subsystem, referred to the hole 
notation. In the Mott insulating ground state, all these
levels are coupled by virtual hoppings resulting in a multiorbital
superexchange interaction operating within the Hilbert space spanned by
$J_{\textrm{eff}}=1/2$ and $J_{\textrm{eff}}=3/2$ states.  

\begin{figure}[b]
\centering \epsfig{file=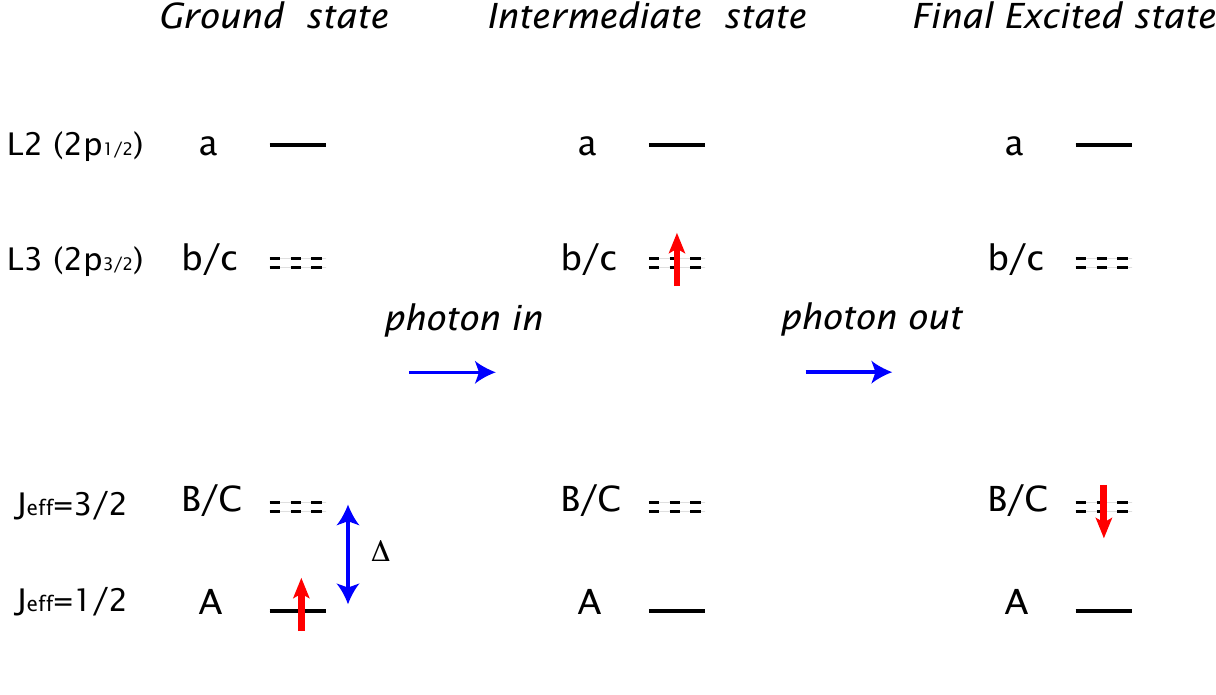, width=7.5 cm}
\renewcommand{\thefigure}{S\arabic{figure}}
\caption{The RIXS process in the hole representation.}
\end{figure}

Figure S1 illustrates the RIXS process in the hole representation. The hole in
the $A$ level is excited to the 2$p$ core levels, also
consisting of three doubles ($a$, $b$, and $c$), by absorbing a photon. Subsequently, the excited intermediate state decays back to the $t_{2g}$ manifold, emitting a photon. The excited hole decaying back to the $A$
level with opposite $\sigma$ corresponds to an effective isospin flip, and the $B$/$C$
level to an effective spin/orbital flip relative to the orbital/spin.    
To describe the RIXS process mathematically, we define operators $X_\sigma$,
$Y_\sigma$ and $Z_\sigma$ for annihilating a hole in the $yz$, $zx$, and $xy$
orbital, respectively. These operators can be expressed in terms of the
$A_\sigma$, $B_\sigma$, and $C_\sigma$ Kramers doublets as 
\begin{eqnarray}
X_\sigma&=&-\frac{\sigma}{\sqrt{2}}(\cos\theta A_{-\sigma}-
\sin\theta B_{-\sigma}+C_{-\sigma}), \\
Y_\sigma&=&\frac{i}{\sqrt{2}}(\cos\theta A_{-\sigma}-
\sin\theta B_{-\sigma}-C_{-\sigma}),\\
Z_\sigma&=&\sin\theta A_{\sigma}+\cos\theta B_{\sigma}.
\end{eqnarray}
Likewise, we define operators for creating a hole in the $2p$ manifold as
\begin{eqnarray}
x_\sigma&=&-\frac{\sigma}{\sqrt{2}}(\cos\phi a_{-\sigma}-
\sin\phi b_{-\sigma}+c_{-\sigma}), \\
y_\sigma&=&\frac{i}{\sqrt{2}}(\cos\phi a_{-\sigma}-
\sin\phi b_{-\sigma}-c_{-\sigma}),\\  
z_\sigma&=&\sin\phi a_{\sigma}+\cos\phi b_{\sigma}. 
\end{eqnarray}
When the spin-orbit coupling is strong and dominates the crystal field effects, as 
realized in Sr$_2$IrO$_4$$^{\textrm S3}$), then 
$\cos\phi=\cos\theta=\sqrt{2/3}$.    

The effective RIXS amplitude $A_{\mathbf q}$ can be defined as  
\begin{equation}
A_{\mathbf q} \propto \langle f \mid \sum_{\mathbf R} e^{i \mathbf q\cdot\mathbf R} 
D^\dag(\epsilon')D(\epsilon)\mid i\rangle,
\end{equation}
where $D$ given by
\begin{equation}
D=\sum_{\sigma}  
(\epsilon_yz^\dag+\epsilon_zy^\dag)_\sigma X_\sigma +
(\epsilon_zx^\dag+\epsilon_xz^\dag)_\sigma Y_\sigma +
(\epsilon_xy^\dag+\epsilon_yx^\dag)_\sigma Z_\sigma 
\end{equation}
is the dipole transition operator for exciting the valence hole into the $p$
core level, a function of incoming photon polarization $\epsilon$. Likewise
$D^\dag$ describes the decay and is a function of the outgoing photon
polarization $\epsilon'$. 

Integrating out the $p$ manifold and using Eqs.1-3, the effective RIXS 
operator $\hat{R}\equiv D^{\dag}(\epsilon')D(\epsilon)$ is obtained. 
This operator annihilates a particle in the $A$ doublet and creates a particle in the $A$ doublet
or in the $B$/$C$ subsystem:
\begin{align}
\hat{R} &= \sum_{\sigma,\sigma'} 
\left(\alpha_{\sigma,\sigma'}^{\phantom{\dagger}} A^{\dagger}_{\sigma'}
+\beta_{\sigma,\sigma'}^{\phantom{\dagger}} B^{\dagger}_{\sigma'}
+\gamma_{\sigma,\sigma'}^{\phantom{\dagger}} C^{\dagger}_{\sigma'}\right)
A^{\phantom{\dagger}}_{\sigma} \nonumber\\
&\equiv \hat{R}_A + \hat{R}_B + \hat{R}_C \;.
\end{align}
The coefficients $\alpha_{\sigma\sigma'}$, $\beta_{\sigma\sigma'}$, and 
$\gamma_{\sigma\sigma'}$ depend on the scattering geometry and determine the
intensities. For the $L_2$ edge, they all vanish in the limit of cubic 
symmetry; shown below are the results for the $L_3$ edge relevant here. 
The first term $\hat{R}_A$ describes processes involving the $A$ doublet 
only and is thus sensitive to magnons. Expressing $A^\dag A$ in terms of 
isospin operators, $\hat{R}_A$ can be recast into the form  
\begin{equation}\label{eq:spinop} 
\hat{R}_A=i \frac{2}{3} (P_x S^x + P_y S^y - P_z S^z),
\end{equation}
where ${\mathbf P}$=$\mathbf{\epsilon'}\times\mathbf{\epsilon}$ and {\bf S}
denotes the isospin within the A doublet. The minus sign in front of the 
$S^z $component reflects the fact that the effective $g$ factor of this doublet 
is opposite to that of real spin because of large orbital moment$^{\textrm S4}$. 
In principle, the $\hat{R}_A$ can be transformed to a scalar product form, 
by a simultaneous sign change of the $S^x, S^y$ components.

Performing similar calculations, the matrix elements for the processes
described by $\hat{R}_B$ and $\hat{R}_C$, which create a spin-orbit exciton
by promoting a hole from the A states to the B/C levels, can be derived as 
\begin{eqnarray}\label{eq:coeffs_bc}
\beta_{\sigma,\sigma} &=& \frac{1}{\sqrt{6}}(Q_3+i\sigma\frac{1}{\sqrt{3}}P_z),\\
\beta_{\sigma,-\sigma} &=& \frac{1}{2\sqrt{2}}
\left[\sigma (T_y-\frac{1}{3}P_y)+i(T_x+\frac{1}{3}P_x)\right],\\
\gamma_{\sigma,\sigma} &=& \frac{1}{\sqrt{6}}(Q_2+i\sigma T_z),\\
\gamma_{\sigma,-\sigma} &=& \frac{1}{2\sqrt{6}}
\left[-\sigma (T_y+P_y)+i(T_x-P_x)\right],
\end{eqnarray}
with
\begin{eqnarray} 
Q_2 &=& \epsilon'_y\epsilon_y - \epsilon'_x\epsilon_x,\\ 
Q_3 &=& \frac{1}{\sqrt{3}}(\vec{\epsilon'}\cdot\vec{\epsilon} -
3\epsilon'_z\epsilon_z),\\
T_x &=& \epsilon'_y\epsilon_z + \epsilon'_z\epsilon_y,\\ 
T_y &=& \epsilon'_x\epsilon_z + \epsilon'_z\epsilon_x,\\ 
T_z &=& \epsilon'_y\epsilon_x + \epsilon'_x\epsilon_y.
\end{eqnarray}

\section{Modeling the spin-orbiton dispersion}

In a leading approximation (neglecting the intensity transfer between the exciton and 
magnons due to a mode-mode coupling), the combined spectral weight of 
$\hat{R}_B+\hat{R}_C$ transitions averaged over the outgoing polarization $\vec{\epsilon'}$ is $q$-independent, and is given by
$\sum_{\sigma,\sigma'}(|\beta_{\sigma,\sigma'}|^2+|\gamma_{\sigma,\sigma'}|^2)
=\frac{2}{9}$.  The mode-mode coupling is expected to bring only weak $q$ 
dependence to the exciton intensity since the related corrections (of the 
order of $J/\Delta$) are small. This agrees well with the experimental fact that the spectral weight of the high-energy mode is nearly momentum independent.

On the other hand, the magnon intensity which is determined by a dynamical 
spin structure factor  strongly depends on $q$, see Eq.~\ref{eq:spinop}.  
For the present experimental setup, the RIXS signal is determined by a 
following combination of the transverse $xx$, $yy$ and longitudinal 
$zz$ components of the isospin susceptibility:
\begin{align}\label{eq:chiA}
I_A({\bf q},\omega)=\frac{1}{9\pi}
(\frac{3}{2}\chi^{xx}+\chi^{yy}+\frac{3}{2}\chi^{zz})^{''}_{{\bf q},\omega}\;,
\end{align}
from which the intensities of single-magnon $I^{(1)}_A({\bf q})$ and 
two-magnon $I^{(2)}_A({\bf q})$ processes can be calculated. (In this 
expression, we neglected weak $q$ dependence of the polarization 
factors). Within a linear spin-wave theory, we find:   
\begin{align}\label{eq:weights_magnon} 
&\quad I^{(1)}_A({\bf q}) =  
\frac{1}{9}\;\frac{5}{8}\sqrt{\frac{A_{\bf q}-B_{\bf q}}{A_{\bf q}+B_{\bf q}}},
\nonumber\\
&\quad I^{(2)}_A({\bf q}) =  
\frac{1}{9}\;\frac{3}{8}\sum_{\bf k}\left(\frac{A_{\bf k}A_{{\bf k}+{\bf q}}
-B_{\bf k}B_{{\bf k}+{\bf q}}}{\omega_{\bf k}\omega_{{\bf k}+{\bf q}}}-1\right)\;,
\end{align}
with $A_{\bf q}$, $B_{\bf q}$ and magnon dispersion $\omega_{\bf q}$ given 
below in Eqs.~\ref{eq:bog_AB} and \ref{eq:bog_omega}.

\begin{figure}
\centering \epsfig{file=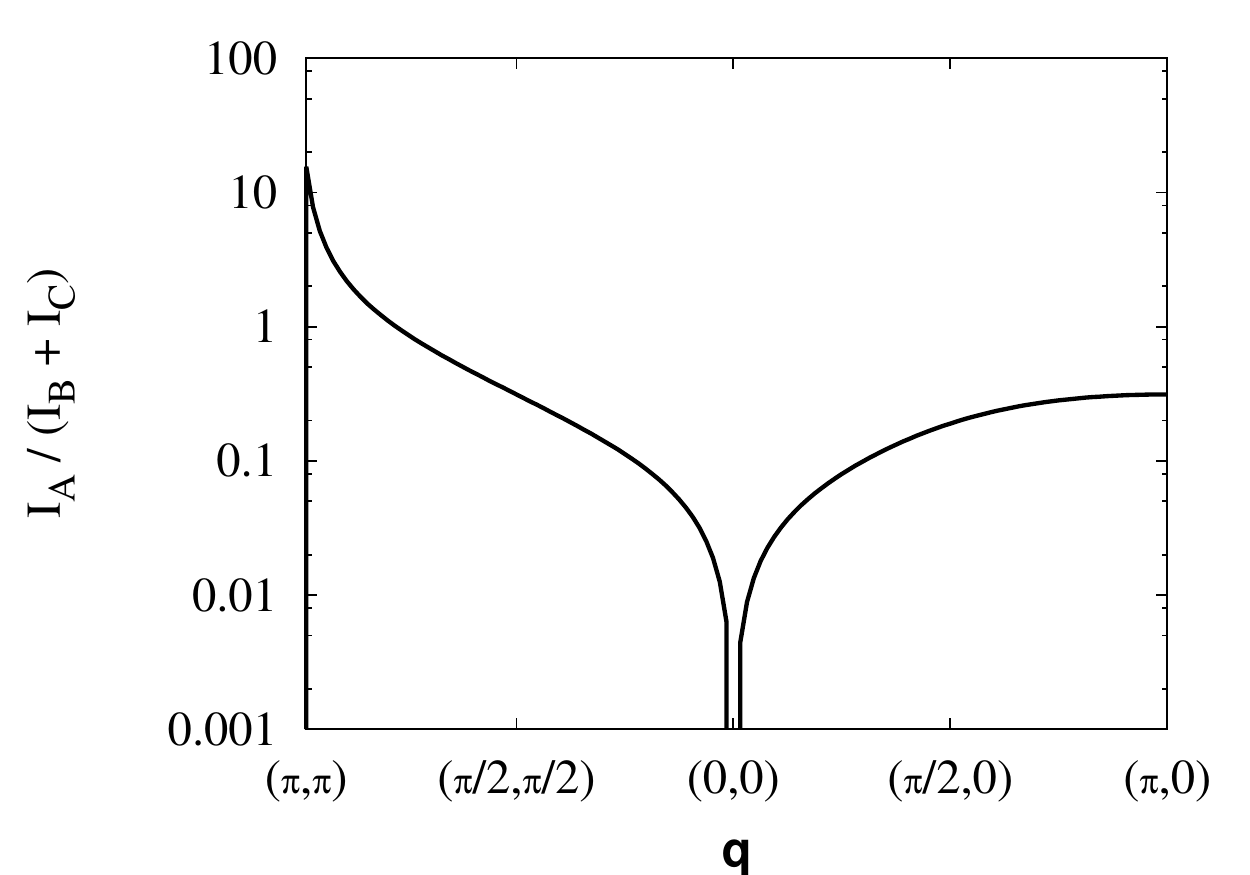,width=7.5 cm}
\renewcommand{\thefigure}{S\arabic{figure}}
\caption{Spectral weight of the magnon excitation divided by the weight 
of the excitonic sector at higher energy from the calculation.}
\label{fig:weights}
\end{figure}

The relative intensity of the magnon and excitonic modes is a 
parameter-free result and provides an important test for the consistency of
the model and the assignment of the observed modes. The ratio of the intensities 
of the magnon and the excitonic high-energy modes is shown in 
Fig.~\ref{fig:weights}. For $q=(\pi,0)$, for instance, theory predicts the 
ratio $\simeq 0.3$ which compares well with the experimental value 
0.26 of the relative intensities of the low- and high-energy modes. 
We now discuss the origin of the dispersive features in the excitonic channel. 
The exciton created by $\hat{R}_B$ and $\hat{R}_C$ delocalizes 
due to virtual hoppings of electrons as depicted in Fig.~3c. 
In essence, this propagation is due to a multiorbital superexchange 
interaction which is responsible for magnetic ordering of the $A$ Kramers 
doublets, but involving here both the ground $J_{\textrm{eff}}$=1/2 state and 
excited $J_{\textrm{eff}}$=3/2 states of Ir ions. 
As can be seen in Fig.~3c, the two holes
(drawn as isospin arrows) are always in different orbitals. (Virtual
processes involving doubly occupied orbitals exist, but do not
contribute to exciton delocalization and hence will not be
considered here.) If we neglect the Hund's rule coupling, this
implies that the spin of the excited electron plays no role, i.e., the
hole-excitation pair can be seen as just a spinless hole, as in the
``usual'' $t$-$J$ model derived from the one-band Hubbard model. The
hole has, however, an orbital flavor: $B$ or $C$.

In the two-dimensional space spanned by these two orbitals, the
effective hopping matrix for the exciton reads
\begin{align}\label{eq:eff_hopp}
W^{\beta\gamma}_{x/y} = -\frac{2t_{AA}}{U'}\left(\begin{array}{cc}
t_{BB}& \pm t_{BC}\\
\pm t_{BC}& t_{CC}
\end{array} \right)\;,
\end{align}
where the inter-orbital element changes sign between the $x$ and $y$ 
directions. One clearly sees the similarity to the superexchange: the effective 
exciton hopping is $\propto \tfrac{t^2}{U'}$, i.e., of the order of $J$ 
instead of $t$. [The factor 2 arises because the second step in 
Fig.~3c can also occur first.]

While the spin of the excited $B$/$C$ state does not enter the
Hamiltonian, the exciton creates a hole in the ordered isospin
background of the $A$ system below the magnetic ordering
temperature. Following Ref.~S5,  this hopping
connected to magnon creation/annihilation can be written as 
\begin{equation}
H_{\textrm{kin}}=\sum_{i,j,\alpha,\beta} W^{\alpha\beta}_{ij} \; X^{\dag}_{i\alpha} 
X^{\phantom{\dagger}}_{j\beta} \; (b^{\dag}_j + b^{\phantom{\dagger}}_i), 
\label{eq:H}
\end{equation}
where we denoted a spin-orbit exciton by $X_{i,\alpha}$. It carries
an index $i$ indicating a lattice site and a quantum number $\alpha$ 
indicating whether it is in the $B$ or $C$ doublets. We see that a hopping 
of $X$ particle is accompanied by a creation $b^{\dag}$ or annihilation $b$ 
of magnon excitations. $W^{\alpha\beta}_{ij}$ is the orbital- and
direction-dependent effective hopping amplitude, see
Eq.~\ref{eq:eff_hopp}. We consider here only the nearest-neighbor hoppings. 

Comparison of our effective Hamiltonian to that of a hole moving in
a quantum antiferromagnet reveals that the two problems are
equivalent, even though we do not describe a real charge moving at an
energy scale of $t$, but exciton motion with a strongly renormalized
hopping matrix $W^{\alpha\beta}=W\tau^{\alpha\beta}$. The energy scale 
$W=2t_{AA}^2/U'$ quantifies the exciton hopping amplitude, and
Eq.~\ref{eq:eff_hopp} becomes
\begin{align}\label{eq:eff_hopp2}
W^{\alpha\beta}_{x/y} = -W\left(\begin{array}{cc}
t_{BB}/t_{AA}& \pm t_{BC}/t_{AA}\\
\pm t_{BC}/t_{AA}& t_{CC}/t_{AA}
\end{array} \right)=
-W\tau^{\alpha\beta}_{x/y}
\end{align}
where $\tau^{\alpha\beta}$ are numbers of the order
of one ($t_{BB}/t_{AA}=\tfrac{5}{4}, t_{BC}/t_{AA}=\pm\tfrac{\sqrt{3}}{4},
t_{CC}/t_{AA}=\tfrac{3}{4}$ in the cubic limit used here; the $\pm$ sign 
depends on $x/y$ bond direction). For nearest-neighbor hopping of the
$J_{\textrm{eff}}=1/2$ states, $t_{AA}\simeq 0.26$~eV is suggested from band
structure calculations$^{\textrm S6-S8}$, and using a representative value of interorbital 
Coulomb interaction $U'\sim 2$ eV, one estimates $W\simeq 60-70$~meV. 

\begin{figure}[tbp]
\centering \epsfig{file=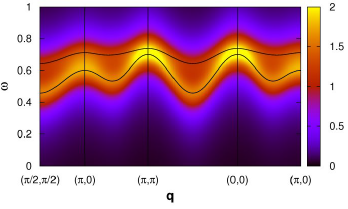,width=8 cm}
\renewcommand{\thefigure}{S\arabic{figure}}
\caption{Dispersion of the hole-excitation pair. Color shading shows
the spectrum convoluted with a Lorentzian broadening with width
$\delta$=100 meV and weighted with their intensity, 
see Eq.~\ref{eq:weights}, the lines show the eigenvalues of
$\Delta$+$\Sigma_{\mathbf k}$, see Eq.~\ref{eq:sigma2}. 
Parameters: $W$=63 meV, $J_1$=60 meV, 
$J_2$=-20 meV, $J_2$=15 meV, and 
$\Delta$ =0.78 eV.} 
\label{fig:Ak}
\end{figure}

Each hopping process creates or destroys magnons. We follow
Ref.~S5 and treat the AF background in the linear spin-wave 
theory. After the Bogoliubov transformation, the correlated hopping of exciton 
(accompanied by a magnon creation/annihilation) reads as follows:
\begin{align}\label{eq:ham_kin}
H_{\textrm{kin}} =-zW \sum_{{\bf k},{\bf q},\alpha,\beta} 
\Bigl[M^{\alpha\beta}_{{\bf k},{\bf q}}
\hdag_{\alpha{\bf k}}\hnod_{\beta{\bf k}-{\bf q}} \anod_{{\bf q}} + h.c. \Bigr]\;.
\end{align}
Except for the orbital index, this is analogous to the $t$-$J$ model for 
a doped hole motion in AF-ordered cuprates. Here, 
$\hdag_{\alpha{\bf k}}$ ($\hnod_{\alpha{\bf k}}$) creates (annihilates) an 
exciton with momentum ${\bf k}$ and orbital flavor $\alpha$, while 
$\adag_{\bf k}$ ($\anod_{\bf k}$) creates (annihilates) a magnon, and
$z=4$ is the coordination number. A momentum- and orbital-dependent 
vertex has been introduced as
\begin{align}\label{eq:vertex}
M^{\alpha\beta}_{{\bf k},{\bf q}}=
|\tau^{\alpha\beta}|(u_{\bf q}\gamma^{\alpha\beta}_{{\bf k}-{\bf q}}
+ v_{\bf q}\gamma^{\alpha\beta}_{{\bf k}})\;,
\end{align}
absorbing the sign change of $t_{BC}$ into the following form-factors:  
\begin{align}
\gamma^{\alpha\beta}_{\bf k}=\frac{1}{2}\left(\begin{array}{cc}
(\cos k_x + \cos k_y)& (\cos k_x - \cos k_y)\\
(\cos k_x - \cos k_y)& (\cos k_x + \cos k_y)
\end{array} \right)\;.
\end{align}
$u_{\bf q}$ and $v_{\bf q}$ are Bogoliubov factors and given by the
relations  
\begin{align}\label{eq:bog_uv}
u_{\bf q} = \frac{1}{\sqrt{2}} \sqrt{\frac{A_{\bf q}}{\omega_{\bf q}}+1}, \quad 
v_{\bf q} = -\frac{\textrm{sgn} B_{\bf q}}{\sqrt{2}}
\sqrt{\frac{A_{\bf q}}{\omega_{\bf q}}-1} 
\end{align}
with
\begin{align}\label{eq:bog_AB}
A_{\bf q} &= 2(J_1-J_2-J_3+J_2\cos q_x\cos q_y)+J_3(\cos 2q_x+\cos 2q_y), 
\nonumber\\ 
B_{\bf q} &= J_1(\cos q_x+\cos q_y), 
\end{align}
and the magnon energy
\begin{align}\label{eq:bog_omega}
\omega_{\bf q} &= \sqrt{A_{\bf q}^2-B_{\bf q}^2}\;.
\end{align}
The parameters $J_1$, $J_2$ and $J_3$ denote nearest-neighbor,
next-nearest neighbor, and third-neighbor couplings of the 
isospins $\vec{S}_A$.

The total Hamiltonian describing the exciton motion in the AF-ordered 
background contains the kinetic energy $H_{\textrm{kin}}$ as well as the
energy of the magnon sector 
$H_{\textrm{mag}}= \sum_{\bf q} \omega_{\bf q}^{\phantom{\dagger}}
\adag_{{\bf q}}\anod_{{\bf q}}$. Since $W^{\alpha\beta}_{x/y}$ is of the order 
of the $J$'s rather than the bare hopping $t$, the
present case is slightly different from the ``usual'' $t$-$J$ model
obtained as a large-$U$ limit of the Hubbard model, where $t \gg J$. The
latter describes a strong coupling of holes to magnons, while the present 
case has a coupling comparable to the magnon energy, $W\sim J$, since both 
parameters originate from the same multiorbital superexchange interaction. 
We consequently use perturbation theory to describe the exciton moving 
in the AF background. Second order perturbation theory leads to an energy 
correction
\begin{align}\label{eq:sigma2}
\Sigma_{\bf k}^{\alpha\beta}=-(zW)^2 \sum_{{\bf q}, \gamma} \frac{
M^{\alpha\gamma}_{{\bf k},{\bf q}}M^{\gamma\beta}_{{\bf k},{\bf q}}} {\omega_{\bf q}}, 
\end{align}
with the vertex $M^{\alpha\beta}_{{\bf k},{\bf q}}$ defined in
Eq.~\ref{eq:vertex}. This sum can be evaluated numerically, and
with parameters $W=63\;\textrm{meV}$ and $\Delta =0.78\;\textrm{eV}$ we 
find the dispersive feature shown in Fig.~\ref{fig:Ak}.

In Fig.~\ref{fig:Ak}, the lines give the two eigenvalues of 
$\Sigma_{\bf k}^{\alpha\beta}$, which determine the excitation energies. The
weight of the poles can be obtained from the prefactors given in
Eq.~\ref{eq:coeffs_bc} and the eigenvectors $\cos \theta_{1(2){\bf k}}, \sin
\theta_{1(2){\bf k}}$ corresponding to the two eigenvalues of 
$\Sigma_{\bf k}^{\alpha\beta}$. The spectral weight of each pole 
(labeled by 1 and 2) is given by 
\begin{align}\label{eq:weights}
I_{1(2)}({\bf k})&=
\sum_{\sigma,\sigma'}\left[|\beta_{\sigma,\sigma'}|^2\cos^2 \theta_{1(2)}+
|\gamma_{\sigma,\sigma'}|^2\sin^2 \theta_{1(2)} \right.\nonumber\\
&\quad\left.+\sin \theta_{1(2)} \cos \theta_{1(2)} 
(\gamma_{\sigma,\sigma'}^*\beta_{\sigma,\sigma'} + 
\gamma_{\sigma,\sigma'}\beta_{\sigma,\sigma'}^*)\right]_{\bf k}\;.
\end{align}
We recall that while the partial intensities $I_1({\bf k})$ and 
$I_2({\bf k})$ are momentum dependent (via the $\cos\theta_{\bf k}$ and 
$\sin\theta_{\bf k}$ functions) affecting thereby the energy/momentum 
intensity map, the total spectral weight $I_1+I_2$ is nearly 
constant ($=\frac{2}{9}$) over an entire Brillouin zone.

\begin{figure}[t]
\centering \epsfig{file=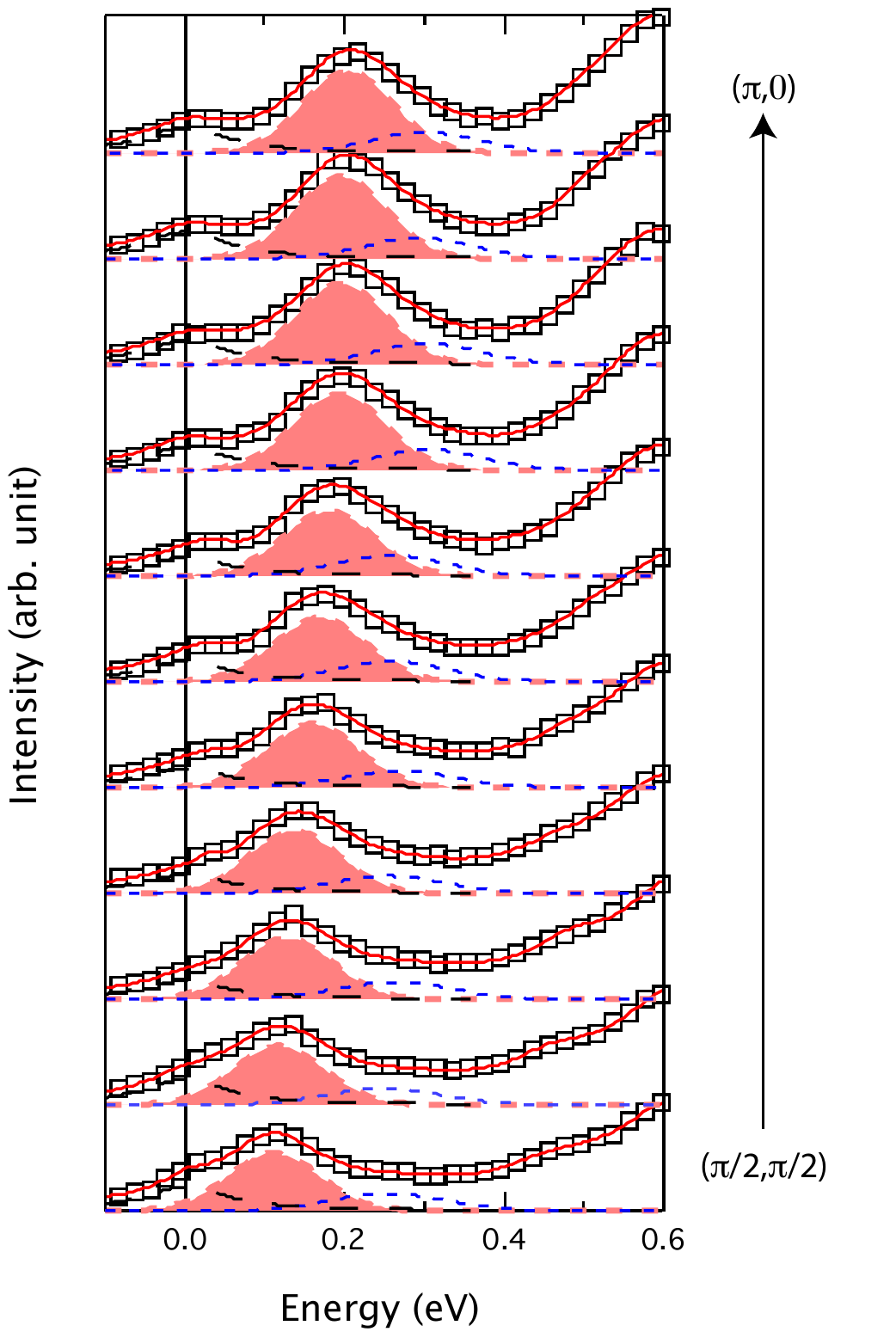, width=7cm}
\renewcommand{\thefigure}{S\arabic{figure}}
\caption{The single magnon mode extracted from peak fitting. The square markers show the experimental data and the red solid lines shows the fit to the data. The red shaded peak corresponds to the single magnon peak. The dashed black and blue line correspond to the elastic and two magnon peaks, respectively.}
\end{figure}

Corrections to the dispersion curves can be calculated within a standard
self-consistent Born approximation (SCBA). However, the basic features (such as 
location of minima at the magnetic Brillouin zone boundary, maxima at $\Gamma$ 
and magnetic Bragg points, overall bandwidth of the order of magnon bandwidth, 
exciton/magnon relative intensities) are expected to remain unchanged since 
these observations are intrinsic to the model. Based on a well-studied case 
of hole motion in quantum antiferromagnets, we may expect that the major
effect of SCBA calculations will be in a substantial broadening of the 
exciton lines.


\section{Extraction of the single magnon dispersion}

Figure S4 shows the low-energy part of the spectra fit with three peaks along the magnetic Brillouin zone. The three peaks account for the elastic, single (shaded peaks), and double magnon peaks. The elastic and single magnon peaks were constrained to have a Gaussian width of 130 meV, which is the instrumental resolution determined from the elastic peak at a nonresonant condition. The high-energy modes above $
\approx$0.4 eV were also fit with Gaussian functions to account for the tail extending to the magnon modes. The double magnon peaks are not clearly resolved, especially near ($\pi$,0), which results in some uncertainties in the single magnon peak positions and amplitudes. However, regardless of the fitting procedure used the single magnon peak position was not affected by more than 5 meV. 

\section {References}
\begin{enumerate}
\renewcommand{\labelenumi}{[S\arabic{enumi}]}
\item L. J. P Ament, G. Khaliullin, and J. van den Brink,  http://arxiv.org/abs/1008.~4862 (2010).
\item G. Jackeli and G. Khaliullin, {Phys. Rev. Lett.} {\bf 102}, 017205 (2009).
\item B. J. Kim {\it et al.}  {Science} {\bf 323}, 1329 (2009).
\item A. Abragam and B. Bleaney, {\it Electron Paramagnetic Resonance of Transition Ions} (Clarendon Press, Oxford, 1970).
\item S. Schmitt-Rink, C. Varma, and A. E. Ruckenstein, 
Phys. Rev. Lett. {\bf 60}, 2793 (1988). 
\item H. Jin, H. Jeong, T. Ozaki, and Yu, {Phys. Rev. B} {\bf 80}, 075112 (2009).
\item H. Watanabe, T. Shirakawa, and S. Yunoki, {Phys. Rev. Lett.} {\bf 105}, 216410 (2010).
\item F. Wang and T. Senthil, {Phys. Rev. Lett.} {\bf 106}, 136402 (2011).

\end{enumerate}